\documentclass[12pt,preprint]{aastex}
\begin{document}

\title{Massive Protoplanetary Disks in the Trapezium Region}

\author{J. A.  Eisner\altaffilmark{1}}
\affil{University of California at Berkeley \\ 
Department of Astronomy \\
601 Campbell Hall \\
Berkeley, CA 94720}
\email{jae@astron.berkeley.edu}

\and 

\author{John M. Carpenter}
\affil{California Institute of Technology \\ 
Department of Astronomy MC 105-24 \\
Pasadena, CA 91125}
\email{jmc@astro.caltech.edu}

\altaffiltext{1}{Miller Fellow}


\slugcomment{Draft of {\bf \today}}
%
%For submission to {\bf ApJ}}

\begin{abstract}
We determine the disk mass distribution around 336 stars in 
the young  ($\sim 1$ Myr) Orion Nebula cluster by imaging a 
$2\rlap{.}'5 \times 2\rlap{.}'5$ region in $\lambda$3 mm continuum emission
with the Owens Valley Millimeter Array.  For this sample of 336 stars,
we observe 3 mm emission above the  3$\sigma$ 
noise level toward ten sources, six of 
which have also been detected optically in silhouette against the bright 
nebular background.  In addition, we detect 20 objects in 
3 mm continuum emission that do not correspond to 
known near-IR cluster members.
Comparisons of our measured fluxes with longer wavelength 
observations enable rough separation of dust emission from thermal
free-free emission, and we find substantial dust emission toward most
objects.  For the sample of ten objects detected at both 3 mm and near-IR
wavelengths, eight exhibit substantial dust emission.  
Excluding the two high-mass 
stars ($\theta^1$OriA and the BN object) and assuming a gas-to-dust ratio of 
100, we estimate circumstellar masses ranging from 0.13 to 0.39 M$_\odot$.
For the cluster members not detected at 3 mm, images of individual objects
are stacked to constrain the mean 3 mm flux of the 
ensemble.  The average flux is detected at the 3$\sigma$ confidence
level, and implies an average disk mass of 0.005
M$_{\odot}$, comparable to the minimum mass solar nebula. 
The percentage of stars in Orion surrounded 
by disks more massive than $\sim 0.1$ M$_{\odot}$ is consistent with the disk 
mass distribution in Taurus, and we argue that massive disks in Orion do not
appear to be truncated through close encounters with high-mass stars.
%This is contrary to previous results for other rich clusters.
Comparison of the average disk mass and number of massive dusty structures
in Orion with similar surveys of the NGC 2024 and IC 348 clusters 
is used to constrain the evolutionary timescales of massive circumstellar 
disks in clustered environments.  
\end{abstract}

\section{Introduction}
Over the last two decades, high resolution millimeter, infrared, and optical 
images have provided evidence for the existence of 
circumstellar disks on scales of $\sim 0.1$--1000 AU
around young stars \citep[e.g.,][]{KS95,DUTREY+96,
PADGETT+99,OW96,EISNER+04}.  Circumstellar disks are the likely birth-sites 
for planetary systems, and determining their ubiquity, properties, and
lifetimes is crucial for constraining the 
timescales and mechanisms of planet formation.
The mass distribution of protoplanetary disks is especially important since
disk mass is related to the mass of planets that may potentially form.
The minimum-mass protosolar nebula, $\sim 0.01$ M$_{\odot}$ 
\citep{WEID+77,HAYASHI81}, 
is an informative benchmark against which to compare disk masses around young
stars, and such comparisons can constrain the number of protoplanetary disks
with the potential to form planetary
systems like our own solar system.  
%The frequency and lifetimes of these 
%massive disks thus constrain formation scenarios for protosolar systems.

While direct imaging at optical through near-IR wavelengths has
provided concrete evidence for a limited number of circumstellar disks
\citep[e.g.,][]{OW96}, and observations of near-IR excess emission have  
shown statistically that most young stars with ages less than a few million 
years still possess inner circumstellar disks 
\citep[e.g.,][]{STROM+89,HLL01b}, 
these studies did not constrain the disk mass distribution.
To probe the bulk of the disk mass,
which resides in cooler, outer disk regions, observations of optically-thin
millimeter emission are needed.

Several investigators have carried out 
comprehensive single-dish mm and sub-mm continuum surveys toward
regions of star formation comprising loose aggregates of stars:
Taurus \citep{BECKWITH+90,OB95,MA01,AW05}, $\rho$ Ophiuchi 
\citep{AM94,NUERNBERGER+98,
MAN98}, Lupus \citep{NCZ97}, Chamaeleon I \citep{HENNING+93}, Serpens
\citep{TS98}, and MBM 12 \citep{ITOH+03,HOGERHEIJDE+02}.  
In Taurus, 37\% of stars appear to possess disks more massive than
$\sim 0.01$ M$_{\odot}$, and the median disk mass is 
$5 \times 10^{-3}$ M$_{\odot}$ \citep{AW05}.  The fraction of massive disks and
the median disk mass is comparable in $\rho$ Ophiuchi \citep{AM94}.

Expanding millimeter continuum surveys to include rich
clusters allows the determination of 
accurate statistics on the frequency and evolution 
of massive disks as a function of both stellar mass and age. Also, since most 
stars in the Galaxy form in rich clusters 
\citep{LADA+91,LSM93,CARPENTER00,LL03},
understanding disk formation 
and evolution in cluster environments is a vital component in our general 
understanding of how stars and planets form.
The main challenge to observing rich clusters at mm-wavelengths 
is that very high angular resolution is required to resolve individual sources
and to distinguish compact disk emission from the more extended emission
of the molecular cloud.  Single-aperture mm-wavelength telescopes
lack sufficient angular resolution, and to date, only three rich
clusters have been observed with mm-wavelength interferometers:
the Orion Nebula cluster \citep{MLL95,BALLY+98,WAW05}, IC 348 
\citep{CARPENTER02}, and NGC 2024 \citep{EC03}. 

The enhanced sensitivity of the most recent observations of Orion has enabled
the detection of several massive ($\ga 0.01$ M$_{\odot}$)
disks \citep{WAW05}, while upper 
limits from other surveys range from $\sim 0.025$--0.17 M$_{\odot}$
\citep{MLL95,BALLY+98}.  Moreover,
extending detection limits by considering as ensembles 
the large numbers ($\ga 100$) of young stars included in the surveys of 
IC 348 and NGC 2024 allowed estimates of 
mean disk masses of $\sim 0.002$ and 0.005 M$_\odot$, respectively
\citep{CARPENTER00,EC03}. While it appears that an average star
aged $\la 1$ Myr still possesses a massive circumstellar disk,
more sensitive observations are necessary to detect directly large numbers
of massive disks at a range of ages, and thereby constrain the mass 
distribution and evolutionary timescales.

Here, we present a new mm-wavelength interferometric survey of
the Orion Nebula cluster (ONC), a young, deeply embedded
stellar cluster that includes the bright, massive Trapezium stars.
The Trapezium region contains hundreds of stars within several
arcminutes, and pre-main-sequence
evolutionary models \citep[e.g.,][]{DM94} fitted to spectroscopic
and/or photometric data indicate that most stars are less than 
approximately one million years old \citep[e.g.,][]{PROSSER+94,HILLENBRAND97}.
Moreover, the standard deviation in the distribution of 
inferred stellar ages is $\la 1$ Myr \citep{HILLENBRAND97}.
Our observations thus provide a snapshot of millimeter emission around
a large number of roughly coeval young stars.  

With the large number of stars in the ONC, we can begin to investigate the
correlation of disk properties with stellar and/or environmental
properties.  Previous investigations of near-IR excess emission
have explored
the dependence of inner disk properties on stellar mass, age, and environment
\citep[e.g.,][]{HILLENBRAND+98,LADA+00}.  
For example, the fraction of stars in Orion exhibiting near-IR excess emission
seems largely independent of stellar age and mass, although there are
indications of a paucity of disks around very massive stars
\citep{HILLENBRAND+98,LADA+00}.  In addition, the inner disk fraction may
decrease at larger cluster radii \citep{HILLENBRAND+98}.
To explore how the properties of the outer disk component 
correlate with such stellar and environmental properties, 
millimeter observations of cool, optically-thin dust emission are necessary.

Our observations represent an improvement over previous work 
because our mosaicked image encompasses more than three times as many
sources as the previous surveys, enabling an improvement of
$\sqrt{3}$ in estimates of frequency and mean mass
of circumstellar disks.  Moreover, the comparable sensitivity of
this survey with previous observations of IC 348 and NGC 2024 allows
a more direct comparison between relatively 
young (NGC 2024: 0.3 Myr; Meyer 1996; Ali 1996), 
intermediate (ONC: 1 Myr; Hillenbrand 1997), and
old (IC 348: 2 Myr; Luhman et al. 1998; Luhman 1999) clusters, 
providing constraints on timescales for disk evolution.  

%In the next section, we describe the ONC region and discuss
%the stellar and protostellar populations.  The observations and results are
%presented in \S \ref{sec:obs} and \S \ref{sec:cont_o2}, 
%and we derive constraints on the
%circumstellar disk masses in \S \ref{sec:disks}.  Finally, we
%compare the results for the ONC to those
%for NGC 2024, IC 348, and Taurus, and discuss the implications for
%disk evolution in rich clusters.

%\section{The Orion Nebula Cluster\label {sec:onc}}

\section{Observations and Data Reduction \label{sec:ovro2_obs}}
\begin{figure}
\plotone{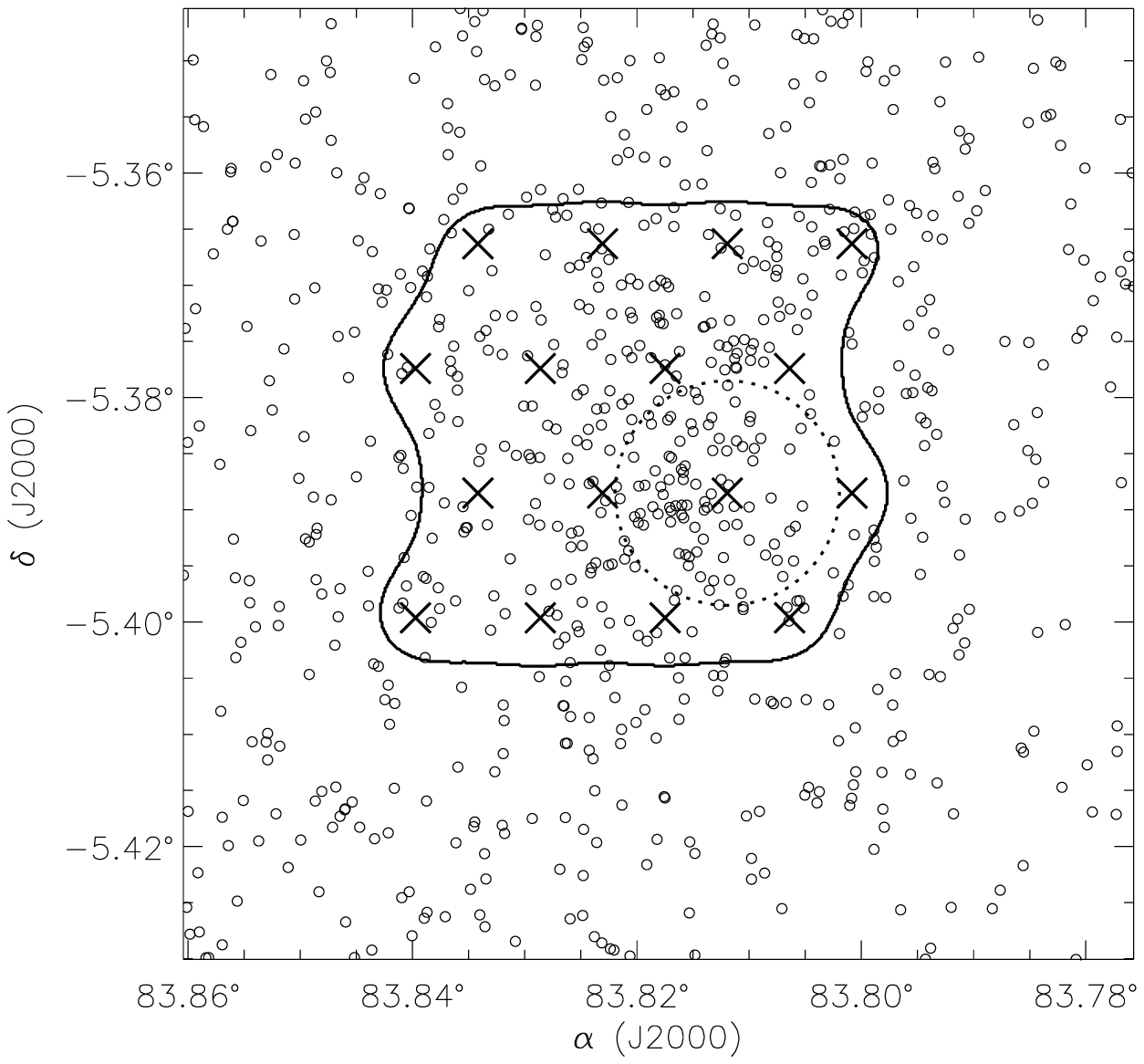}
\caption[Pointing positions for the ONC OVRO mosaic]
{Pointing positions for the OVRO mosaic (``X'' symbols), 
plotted over the positions of $K$-band sources in the ONC (open circles).
These $K$-band source positions are from \citet{HC00}, and have been 
registered to the 2MASS astrometric grid.
The unit gain contour of the mosaic (solid curve) and the FWHM of the
OVRO primary beam at 100 GHz (dotted circle) are also indicated.
\label{fig:pointings_o2}}
\end{figure}

We mosaicked a $2\rlap{.}'5 \times 2\rlap{.}'5$ region toward the
ONC in $\lambda$3 mm continuum with the OVRO millimeter array
between August, 2003 and March, 2004.  Continuum data were recorded 
using the new COBRA correlator, providing a total of 8 GHz of bandwidth
centered at 100 GHz.  Two different
configurations of the 6-element array provided baselines between
35 and 240 meters.   As shown in Figure \ref{fig:pointings_o2},
the mosaic consists of sixteen pointing centers.  For observations in
a given night, the mosaic was observed in its entirety once or twice
(depending on the length of the track), with equal integration time
(and hence equal sensitivity) for each pointing position.

We calibrated the amplitudes and phases of the data with the
blazar J0530+135: 
$(\alpha,\delta)_{\rm J2000} =(5^{\rm h}30^{\rm m}56\rlap{.}^{\rm s}4, 
+13^{\circ}31'55\rlap{.}''2)$.  Three minute observations of J0530+135
were interleaved with sixteen minute integrations on the target mosaic.
We estimated the flux for J0530+135
using Neptune and Uranus as flux calibrators.
%, and 3C84 and 3C273 as secondary calibrators.
Since we obtained data over an eight month time-span, and 
J0530+135 is variable, we estimated the flux for each array configuration.
For observations in the less extended ``E'' configuration (spanning
August-September 2003), we calculated a mean flux of 2.91 Jy based on 
four measurements, and the RMS in the flux measurements was 0.21 Jy.  
For observations in the ``H'' configuration (spanning
December, 2003-March, 2004), we determined a flux of $2.08$ Jy and an RMS of
0.14 Jy using eight measurements. 
All data calibration was performed using a suite of IDL routines
developed for the MIR software package.

We mosaicked the sixteen individual pointings into a single image, robustly
weighted the data (using a robust parameter of 0.5), then
de-convolved and CLEANed the mosaic using the MIRIAD package \citep{STW95}.  
Since we are primarily interested in compact disk emission, we 
eliminated $uv$ spacings shorter than 35 k$\lambda$ in order to avoid 
contamination from bright extended emission.  The eliminated spacings
correspond to size scales larger than $\sim 6''$.  The cutoff  value was chosen
to minimize the RMS background noise in the CLEANed image.
We note that previous analyses of the ONC have also 
eliminated data with uv spacings $<35$ k$\lambda$ \citep{FELLI+93a,BALLY+98}
to filter out the extended emission in the region.

\begin{figure}
\plotone{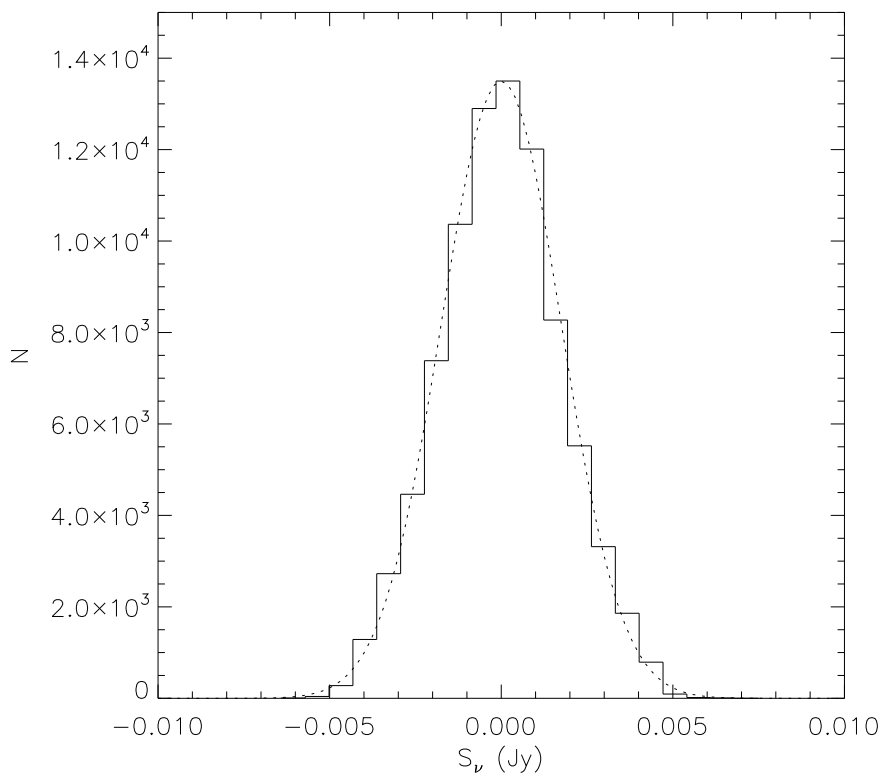}
\caption[Noise distribution in the ONC OVRO mosaic]
{Distribution of 3 mm continuum fluxes for all pixels within the unit gain
contour of a residual image with bright point sources removed using CLEAN
(solid line), and the frequency distribution expected for Gaussian
noise with a mean of zero and a standard deviation of 1.75 mJy (dotted line). 
While the noise varies across the mosaic from 0.88 to 2.34 mJy, as
computed in $0\rlap{.}'5 \times 0\rlap{.}'5$ sub-regions 
(\S \ref{sec:ovro2_obs}), the average RMS noise is 1.75 mJy.
\label{fig:noise}}
\end{figure}

The mosaic produced from our robust-weighted data with $r_{uv} > 35$ k$\lambda$
is shown in Figure \ref{fig:map_o2}.  The unit gain region of the mosaic 
encompasses a $2\rlap{.}'5 \times 2\rlap{.}'5$ area, 
as indicated by the solid curve.
The FWHM of the synthesized beam is $1\rlap{.}''9 \times
1\rlap{.}''5$.   The RMS varies across the mosaic, because
of varying amounts of sidelobe emission from point sources and extended 
emission that were not removed by CLEAN.  
We calculate the RMS of the image in $0\rlap{.}'5 \times  
0\rlap{.}'5$ sub-regions, and find values ranging from 0.88 mJy to 2.34 mJy.  
Despite these variations in RMS, the noise across the mosaic is 
largely Gaussian and the ``effective'' noise for the mosaic is 1.75 mJy
(Figure \ref{fig:noise}).

\section{Results and Analysis \label{sec:cont_o2}}

\subsection{Detected Sources \label{sec:detections}}
\begin{figure}
\plotone{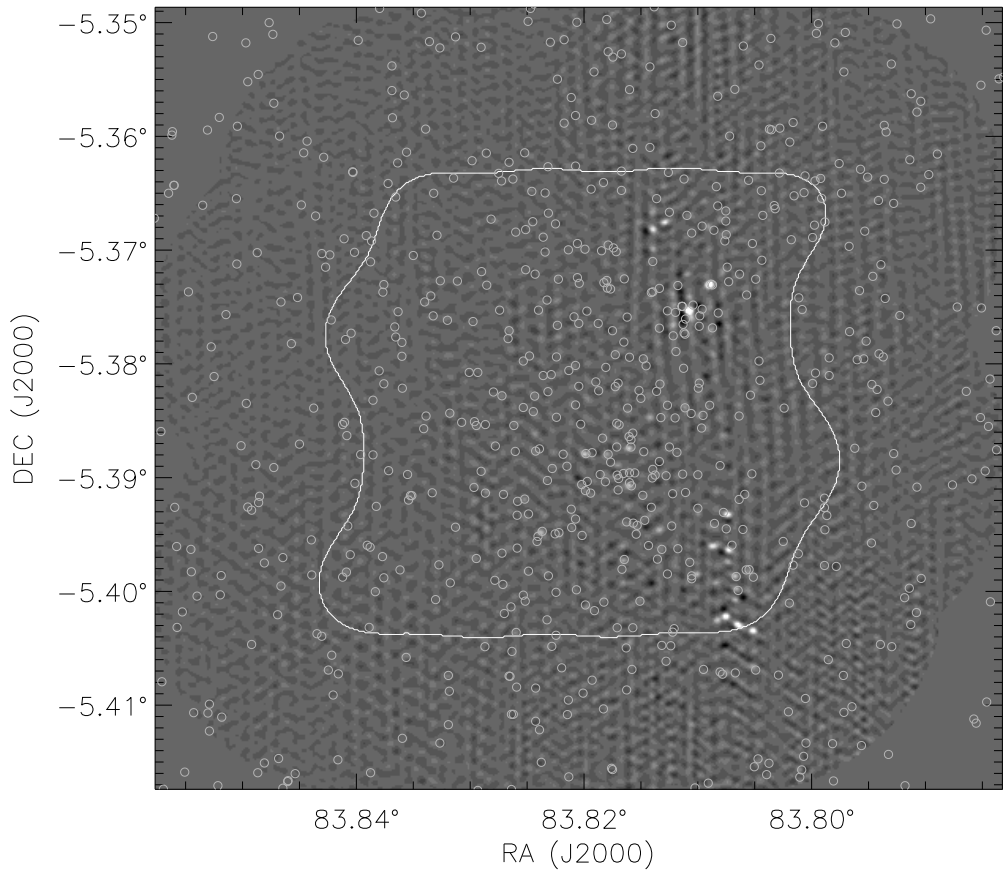}
\caption[$\lambda$3 mm continuum mosaic of the ONC]
{The Orion Nebula cluster, imaged in $\lambda$3 mm
continuum with the Owens Valley Millimeter Array (greyscale).  Only data
observed on long baselines ($r_{uv}>35$ k$\lambda$) were
used to create this image, and the
FWHM of the synthesized beam is $1\rlap{.}''9 \times 1\rlap{.}''5$.
The unit gain region of the mosaic encompasses a 
$2\rlap{.}'5 \times 2\rlap{.}'5$ area, as indicated by the solid contour, and
after CLEANing of the bright point sources apparent in this image,
the average RMS computed for the residuals within the unit gain contour is 
$\sim 1.75$ mJy. Open circles indicate the positions of $K$-band sources from
\citet{HC00}.
\label{fig:map_o2}}
\end{figure}

When searching for detections, we begin by considering only the positions
of known near-IR cluster members \citep{HC00} that lie within the unit gain 
contour of our OVRO mosaic (Figure \ref{fig:map_o2}).
For these 336 pre-determined positions, $\sim 0.4$ sources are expected to show
emission above the 3$\sigma$ level from Gaussian noise, and we therefore
employ a 3$\sigma$ detection threshold, where
$\sigma$ is the noise determined locally in $0\rlap{.}'5 \times  
0\rlap{.}'5$ sub-regions.  

\begin{figure}
\plotone{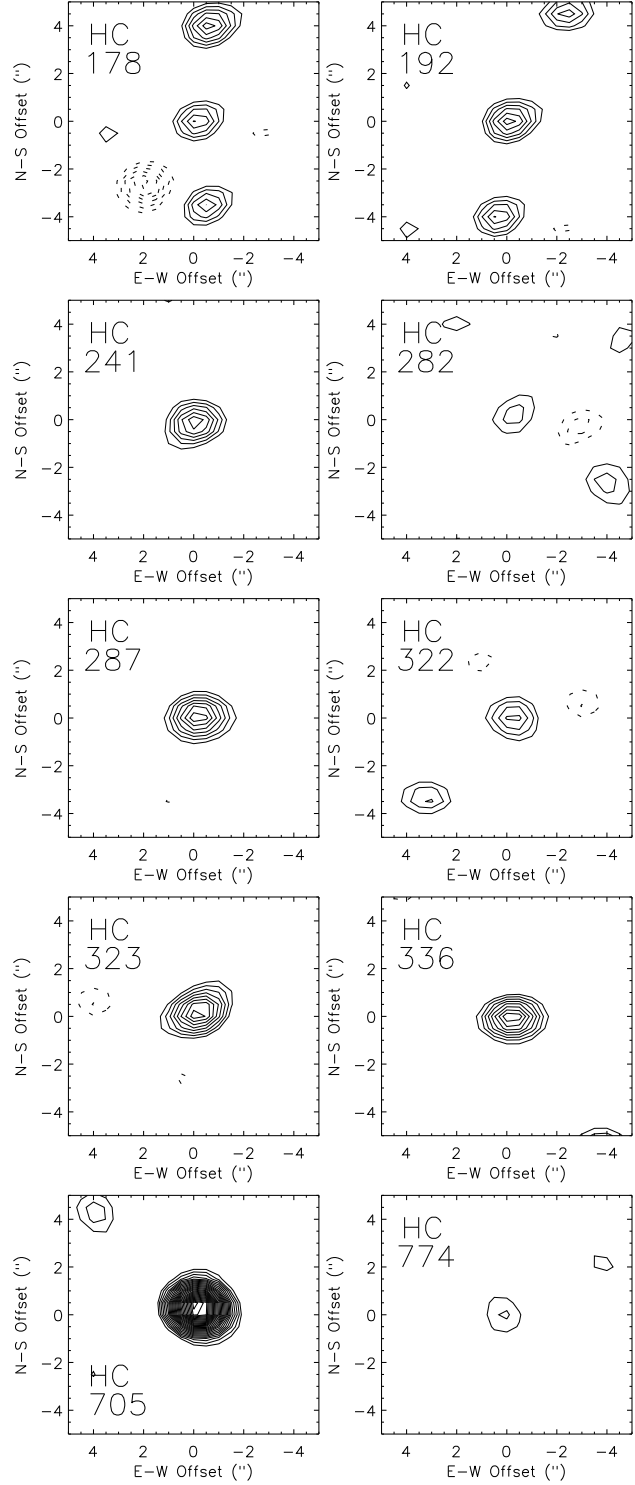}
\caption{Contour images of sources detected at the $\ge 3\sigma$ level
in 3 mm continuum emission
that correspond to previously known near-IR objects.  Contour increments are
1$\sigma$, beginning at $\pm 3 \sigma$, where $\sigma$ is determined 
locally for
each object (see \S \ref{sec:ovro2_obs}). Solid contours represent positive
emission, and dotted contours trace negative features.  HC 705 is the BN
object, and HC 336 is $\theta^1$OriA.
\label{fig:detections}}
\end{figure}

In order to determine whether 3 mm continuum emission is coincident with
near-IR source positions, we estimate relative positional uncertainties
from the centroiding uncertainty for millimeter sources 
($\sim 0.5 \theta_{\rm beam}/$signal-to-noise $\approx 0\rlap{.}''3$) 
and the uncertainty in the near-IR source positions
($\sim 0\rlap{.}''3$).  The 1$\sigma$ relative astrometric uncertainty
is $\sim 0\rlap{.}''4$, and only objects for which the millimeter
emission lies within $0\rlap{.}''4$ of the $K$-band stellar position are
claimed as coincident.  We detect 3 mm continuum emission above the 3$\sigma$ 
level toward 10 near-IR detected objects (Table \ref{tab:detections}).  
The 3 mm emission for these detections appears point-like, with FWHM
comparable to that of the synthesized beam.
OVRO 3mm continuum images of detected sources coincident with
near-IR cluster member positions are displayed in Figure \ref{fig:detections}. 

One of the sources detected at both near-IR and 3 mm wavelengths 
is the BN object \citep{BN67}, and six others correspond with
optically- and radio-detected proplyds \citep{OWH93,FELLI+93a,FELLI+93b}.
Although Gaussian noise is expected to produce
less than one false detection,
there may still be chance coincidences with sidelobe artifacts in regions 
containing bright, nearby
3 mm emission sources.  Thus, HC 178 and HC 192, which occur in crowded
areas displaying bright positive and negative emission 
(Figure \ref{fig:detections}), should be regarded with some caution.

We also search for detections in the mosaic toward positions
that do not correspond with known ONC cluster members.
For the entire mosaic, which
contains substantially more pixels than those considered for near-IR detected
objects, a 5$\sigma$ detection limit is employed to ensure that we are
not ``detecting'' noise spikes in the image: less than one pixel in the
entire image should exhibit a noise spike greater than 5$\sigma$.
Twenty additional objects, without near-IR counterparts, are detected in 
3 mm continuum emission above the 5$\sigma$ level (Table \ref{tab:detections}).
Since many of these detections occur in crowded regions (Figure
\ref{fig:map_o2}), listed fluxes for these objects may be contaminated by
sidelobe emission from other nearby sources.

\subsection{Dust and Free-Free Emission \label{sec:ff}}
Free-free emission arises in hot ionized gas, and in the ONC
such conditions may exist either in HII regions around high-mass stars
\citep[e.g.,][]{GMR87} or in the outer regions of disks or envelopes that are 
irradiated by the hot Trapezium stars \citep[e.g.,][]{OWH93}.
Our 3 mm fluxes may thus trace some free-free emission from hot gas 
in addition to thermal continuum emission from cool dust.  
Because the spectral shape of free-free radiation differs
from that of thermal dust emission, comparing
our 3 mm measurements with longer-wavelength data enables us to
distinguish these components.  Moreover, radio wavelength observations 
have sufficient sensitivity \citep[$\la 0.3$ mJy;][]{FELLI+93b} to detect all 
of the sources detected in our 3 mm mosaic, if the 3 mm flux traces free-free
emission.

We model our observed 3 mm continuum fluxes including both free-free and
dust emission.  The free-free emission is parameterized by a turnover
frequency, $\nu_{\rm turn}$, which demarcates where the radiation switches
from optically-thick to optically-thin, and by the flux at this frequency,
$F_{\rm \nu ,turn}$:
\begin{equation}
F_{\rm \nu, ff} = \cases{F_{\rm \nu, turn} (\nu / \nu_{\rm turn})^{-0.1} &
if $\nu \ge \nu_{\rm turn}$ \cr
F_{\rm \nu , turn} \left({\nu} / {\nu_{\rm turn}}\right)^{2} &
if $\nu \le \nu_{\rm turn}$}.
\label{eq:ff}
\end{equation}
Emission from cool dust is added to this free-free emission to obtain
a model of the observed flux.  We assume that 
\begin{equation}
F_{\rm \nu, dust} = F_{\rm 3 mm, dust} (\nu / 100 {\rm \: GHz})^{(2+\beta)}
= F_{\rm 3 mm, dust} (\nu / 100 {\rm \: GHz})^{3},
\label{eq:dust}
\end{equation}
for $\beta=1$ \citep[e.g.,][]{BECKWITH+90}.  

\begin{figure}
\plotone{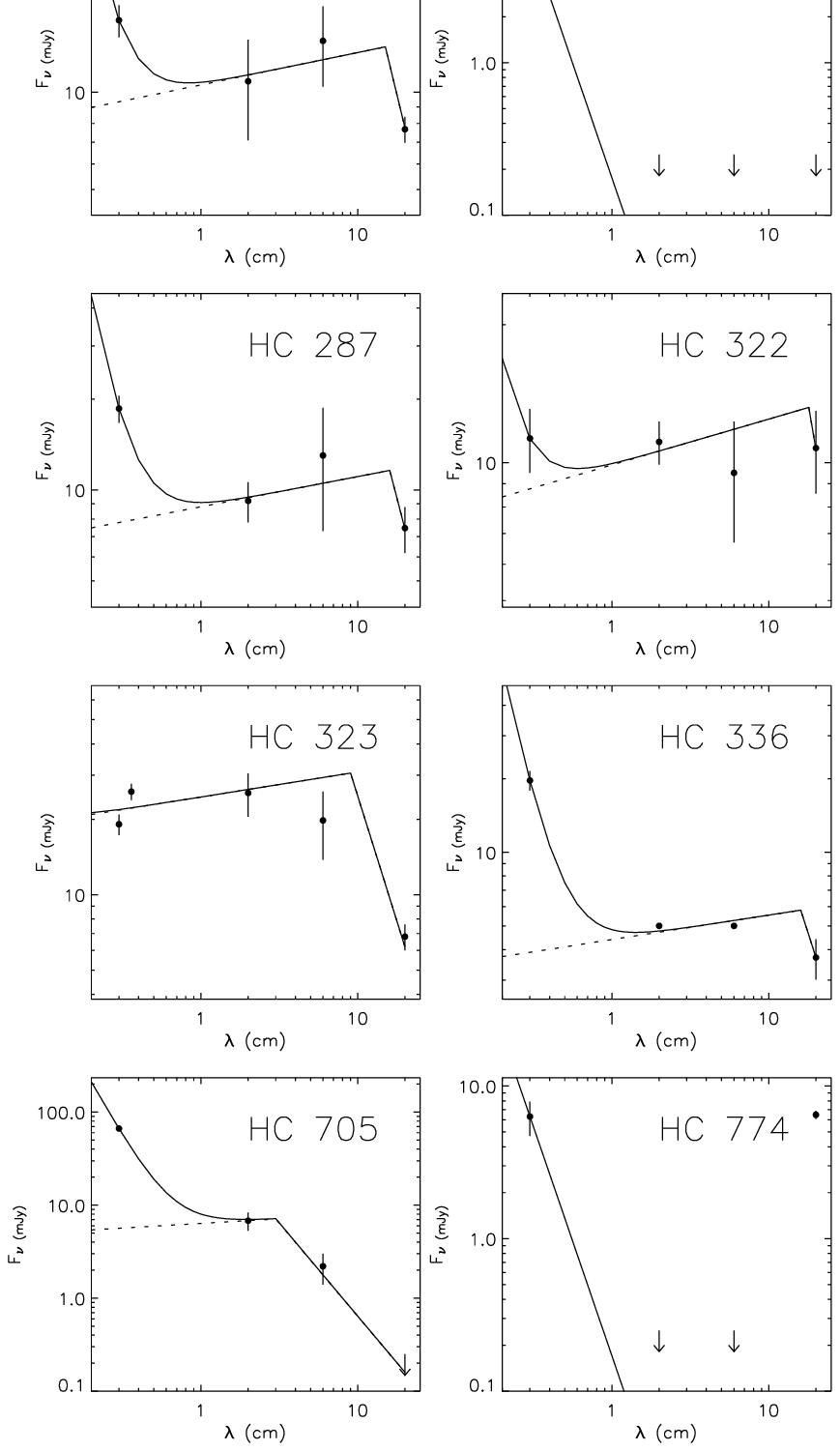}
\caption{Long-wavelength fluxes for our sample (points), along with 
best-fit models including free-free and thermal dust emission.
The free-free emission component is indicated by a dotted line, and
the solid-line represents the combined free-free and thermal dust
emission.  The free-free flux density
is proportional to $\nu^{-0.1}$ for optically-thin
emission and $\nu^2$ for optically-thick; the emission is
thus parameterized by the flux at a single wavelength and a turnover
frequency.  The emission from cool dust is proportional to $\nu^{2+\beta}
\sim \nu^3$ for optically-thin emission with $\beta=1$.
In the case of HC 774, we ignored the 20 cm flux measurement
when fitting the model, since we believe this flux probably arises
from a different emission mechanism.  The 2 and 6 cm emission from 
HC 336 is highly variable \citep[from $\sim 2-80$ mJy;][]{FELLI+93b},
and thus the model fit is uncertain for this object.
\label{fig:seds}}
\end{figure}

We estimate the relative contributions of dust and free-free emission
by fitting this model,  $F_{\nu} = F_{\rm \nu, ff}+F_{\rm \nu, dust}$, to our 
measured 3 mm fluxes and 3.5 mm, 2 cm, 6 cm, and 20 cm fluxes 
from the literature \citep{MLL95,FELLI+93a,FELLI+93b}.  
Given the noise level of the centimeter 
observations \citep[$\la 0.3$ mJy;][]{FELLI+93a,FELLI+93b} and our measured
3 mm fluxes ($\ga 10$ mJy), for objects undetected at centimeter wavelengths,
less than $3\%$ of the 3 mm emission can be free-free.  For
simplicity, we attribute 100\% of the 3 mm flux to dust emission for these
objects.  For objects detected at centimeter wavelengths, we fit the 
dust+free-free model to the $\ge 4$ flux measurements
for each source, and thus we are able to determine the three free parameters
of the model, $\nu_{\rm turn}$, $F_{\rm \nu, turn}$, and $F_{\rm 3mm, dust}$.
Long-wavelength fluxes and models are plotted in Figure \ref{fig:seds}, and the
fluxes due to thermal dust emission 
are listed in column 5 of Table \ref{tab:detections}.  Uncertainties for these
dust fluxes are given by the 1$\sigma$ uncertainties of the model fits.

Three of the objects in Table \ref{tab:detections} that are also seen at
near-IR wavelengths
(HC 178, HC 192, and HC 282) were 
undetected at longer wavelengths, indicating that the 3 mm emission probably
traces dust.  Moreover, although HC 774 is detected at 20 cm, it is
undetected at either 2 or 6 cm, indicating that the long-wavelength measurement
probably traces some non-thermal, variable emission mechanism, while the
OVRO flux likely arises from cool dust emission.
For the remaining 6 sources, which were also detected in
the 2-20 cm surveys of \citet{FELLI+93a,FELLI+93b}, 
at least some components of the measured 3 mm fluxes are likely 
due to free-free emission.  While for most objects some of the observed
flux probably arises from dust emission, for HC 322 and HC 323, 100\% of the 
3 mm flux may be due to free-free emission.  Observations at 850 $\mu$m
showed that even at shorter wavelengths, 100\% of the emission from HC 323
may be free-free, while HC 322 probably has some emission from cool dust
\citep{WAW05}.  

Because HC 336 ($\theta^1$OriA) is highly variable at centimeter 
wavelengths \citep{FELLI+93b}, the flux due to dust listed
in Table \ref{tab:detections} is uncertain.  Previous observations at
3.5 mm wavelength did not detect this object above an RMS noise level of
$\sim 2$ mJy \citep{MLL95}; our measured flux of $\sim 20$ mJy indicates that 
this object is also highly variable at millimeter wavelengths, arguing for 
non-thermal emission.  Thus, the flux component attributed to dust in
Table \ref{tab:detections} is an upper limit.

Among the sources detected at 3 mm but not at near-IR wavelengths, only
one is detected at wavelengths $\ge 2$ cm, IRc2.  For this object, only
$\sim 1\%$ of the flux can be attributed to free-free emission.  
Two other objects (MM8 and MM14) are detected at 1.3 cm wavelength
with flux densities $\la 1$ mJy \citep{ZAPATA+04}; the measured 3 mm
flux densities are $\ga 20$ times higher (Table \ref{tab:detections}).
Thus, it appears that free-free emission does not contribute substantially
to the 3 mm flux from objects without near-IR counterparts;
the lack of free-free emission may indicate that these sources
are very embedded in the cloud, shielded from ionizing radiation.

In our calculation of the relative contributions to the 3 mm flux of 
free-free and dust emission, we assumed that $\beta=1$.  In fact,
$\beta$ may be smaller if grains are substantially larger in protoplanetary
disks than in the interstellar medium \citep[e.g.,][]{MN93,AW05}.  However,
the assumed value of $\beta$ is not crucial to our calculations.  If we
instead assume that $\beta=0$, the derived component of the 3 mm flux due
to dust changes by only a few percent or less.

\subsection{Circumstellar Masses \label{sec:masses}}
The mass of circumstellar material (dust + gas, assuming a standard ISM
gas to dust mass ratio of 100) is related to the component of the 3 mm
continuum flux due to dust emission. 
Assuming the emission is optically-thin, and 
following \citet{HILDEBRAND83},
\begin{equation}
M_{\rm circumstellar} = \frac{S_{\rm \nu,dust} d^2}
{\kappa_{\nu} B_{\nu}(T_{\rm dust})}.
\label{eq:mass_o2}
\end{equation}
Here, $\nu$ is the observed frequency,
$S_{\rm \nu,dust}$ is the observed flux due to cool dust, $d$ is the distance 
to the source,
$\kappa_{\nu} = \kappa_0 (\nu / \nu_0)^{\beta}$ is the mass opacity,
$T_{\rm dust}$ is the dust temperature, and $B_{\nu}$ is the Planck function. 
We assume $d=480$ pc \citep{GENZEL+81}, 
$\kappa_0=0.02$ cm$^{2}$ g$^{-1}$ at 1300 
$\mu$m, $\beta=1.0$ \citep{HILDEBRAND83,BECKWITH+90}, and $T_{\rm dust} = 20$ K
(based on the average dust temperature inferred for Taurus; Andrews \&
Williams 2005; see also the discussion in Carpenter 2002; 
Williams et al. 2005).  While it is possible that external heating by the 
massive Trapezium stars will lead to higher disk temperatures than in Taurus,
we nevertheless assume
that $T_{\rm dust}=20$ K for consistency with previous studies.
Uncertainties in the assumed values of $T_{\rm dust}$ and $\kappa$ imply 
that the derived masses are uncertain (in an absolute sense)
by at least a factor of $3$ \citep[e.g.,][]{POLLACK+94}.
With our assumed values for these parameters, the
masses of detected sources in our OVRO mosaic (Table \ref{tab:detections})
range from 0.13 to 1.45 M$_{\odot}$.  

Since some of the objects in Table \ref{tab:detections} 
may be massive stars, the millimeter flux may contain contributions 
from dust substantially hotter than the assumed 20 K.  
The high luminosity of the BN object
($\sim 2500$ L$_{\odot}$) implies a spectral type of B3-B4 \citep{GBW98},
and the luminosity of $\theta^1$OriA implies a spectral type of O7 
\citep{WH77}. In addition, IRc2 is probably a massive star, and the
high luminosity of nearby ``source I'' provides additional dust heating
\citep{GBW98}.  For these sources,
the computed dust masses are over-estimated, since smaller masses of
hotter dust can produce the observed 3 mm emission (for example,
if $T_{\rm dust}=50$ K then the computed dust masses would be 
lower by a factor of 3.3).
The circumstellar masses listed in Table \ref{tab:detections}
for BN (HC 705), $\theta^1$OriA (HC 336), and IRc2 (MM 3) 
should therefore be treated as upper limits.
Other detected objects do not have
known stellar masses, but are likely to be low-mass stars 
(see \S \ref{sec:nondet}).
Thus, the assumptions in Equation \ref{eq:mass_o2} are
plausible for most of the objects in Table \ref{tab:detections}.

\subsection{Circumstellar Geometry \label{sec:geometry}}
Our OVRO observations alone do not have the angular resolution or
kinematic information necessary to determine whether the
circumstellar material is distributed in disks or envelopes,
or combinations of the two.   Where 3 mm emission is detected
toward known near-IR cluster members, the fact that the near-IR light
is visible despite the high extinctions ($A_{\rm V} \ga 300$) that would arise 
for dust masses inferred from the 3 mm fluxes (for spherically distributed 
material) implies that the dust lies in flattened, disk-like 
distributions \citep[see also, e.g.,][]{BECKWITH+90,EC03}.  There are,
however, exceptions to this argument.  

Massive stars may be surrounded by hotter dust than low-mass stars, and
for these objects smaller dust masses (with smaller associated extinction of 
near-IR light) can produce the observed emission.  Furthermore, the 
high extinctions \citep[$A_{\rm V}\sim 20-60$ mags;][]{GBW98}
observed toward the high-mass stars detected at 3 mm
could be consistent with spherically
distributed material.  Thus, we cannot 
necessarily infer a flattened distribution of dust around $\theta^1$OriA
or BN based on the observed 3 mm fluxes.  
However, most of the stars in our observations have masses 
$\la 1$ M$_\odot$ \citep[e.g.,][]{HC00}, and thus the extinction 
argument generally applies, suggesting that most objects detected in both 
millimeter and near-IR emission trace flattened structures.  

Our claim that objects detected in both 3 mm and near-IR emission are
flattened structures is bolstered by the fact that six
of the detected sources correspond to proplyds observed by \citet{OWH93},
and some of these show flattened disk-like morphologies \citep{BALLY+98a}.
This is particularly striking for HC 774, where optical images
show a flattened ring-like structure surrounding a central, extincted star
\citep{BALLY+98a}.  However, we can not be sure that flattened structures 
observed around these stars are
geometrically thin disks in Keplerian rotation, as opposed to flattened
disk+envelope structures such as those inferred around some Class I objects
\citep[e.g.,][]{EISNER+05b}.

\subsection{Ensemble of Non-Detected Sources \label{sec:nondet}}
Only 3\% of the known near-IR cluster members in the Trapezium region
are detected in 3 mm continuum emission, and thus the vast majority of objects
are undetected in our observations.  In order to examine the flux 
distribution for ``typical'' low-mass stars in the ONC, we consider the
distribution of 3 mm fluxes observed for the ensemble of the
remaining 326 undetected near-IR cluster members; the large number
of sources allows an enhanced sensitivity for the ensemble compared to
that for an individual object.  

While spectra are unavailable for most of the stars in the central region
of the cluster (where we are observing), for the $\sim 125$
objects within the unit gain contour where 
spectra have been obtained, $\sim 85\%$ have stellar masses $\le 1$ 
M$_{\odot}$ \citep{HILLENBRAND97,LUHMAN+00}.  Furthermore, near-infrared
photometric studies of all sources within the unit gain contour confirm that
most stars have masses between  0.1 and $1$ M$_{\odot}$ \citep{HC00}.

\begin{figure}
\plottwo{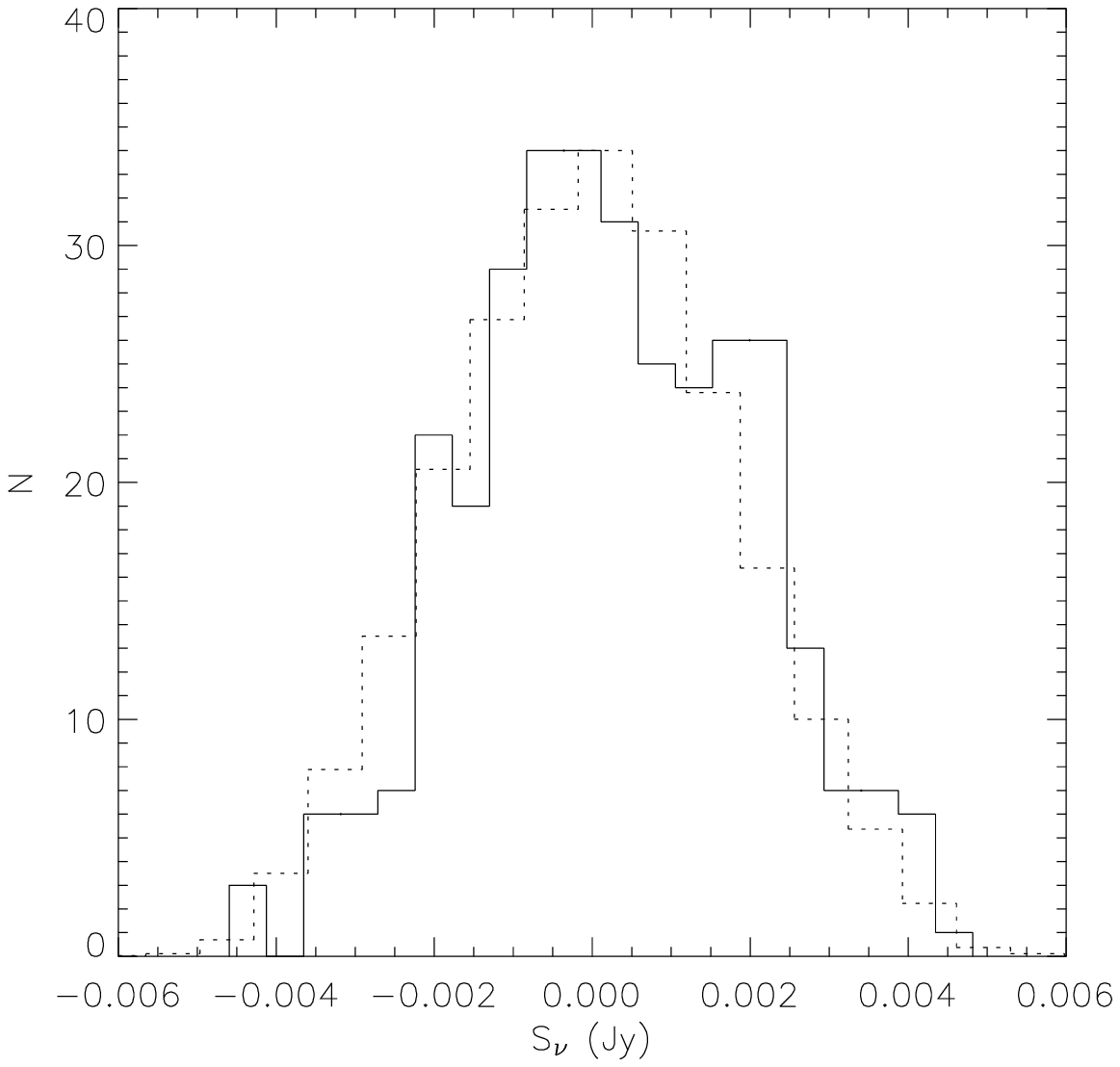}{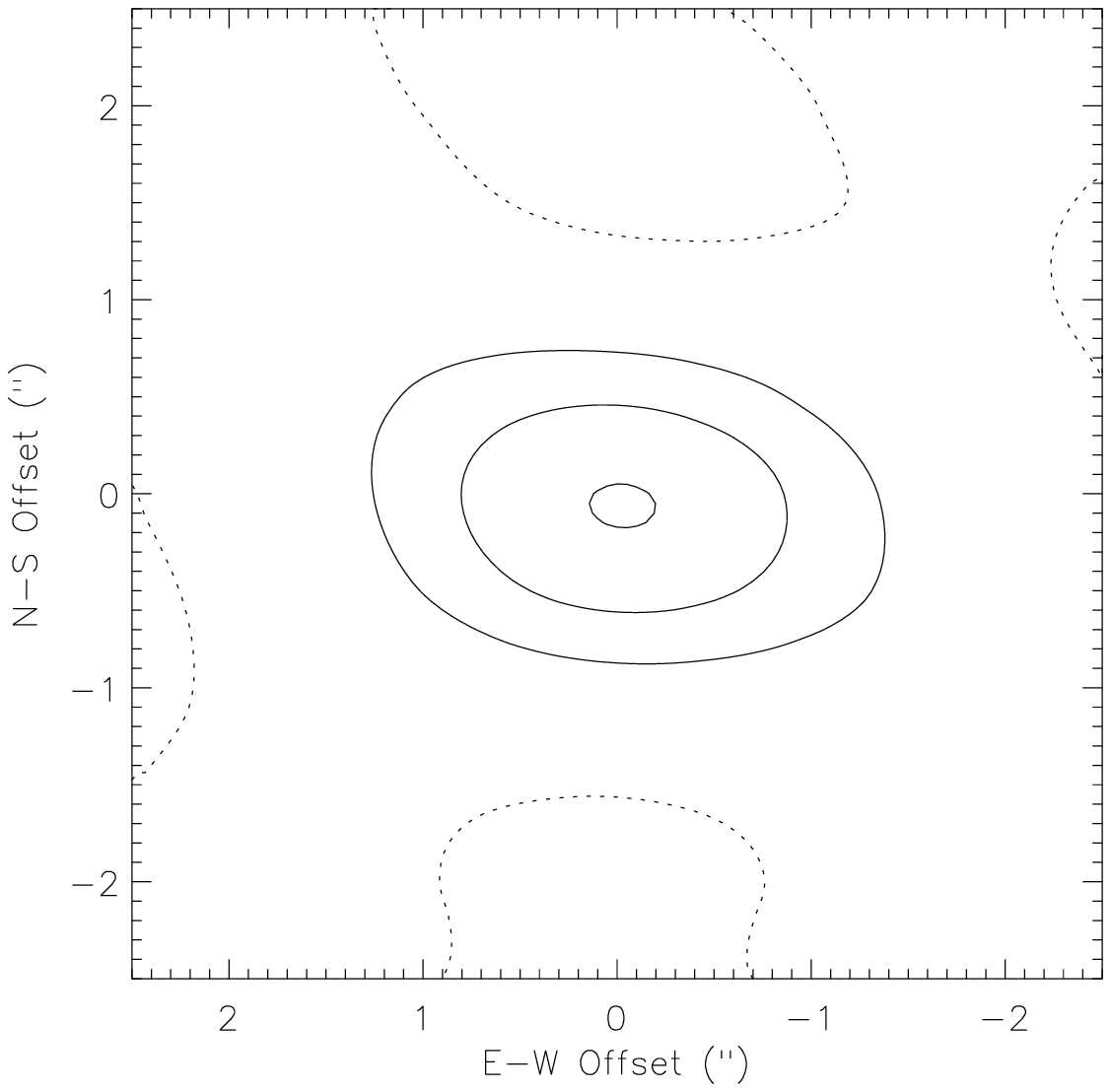}
\caption[3 mm flux distribution and average image for near-IR cluster members]
{(left) Histogram of 3 mm continuum fluxes observed toward positions
of 326 near-IR cluster members in the ONC (solid histogram) and the fluxes
observed toward the remaining positions within the unit gain contour
(dotted histogram).  The flux distribution for near-IR sources is biased
to positive values with respect to the noise distribution. (right) 
Average image, obtained by stacking the 3 mm continuum emission observed
toward each of the 326 low-mass near-IR sources.  Contour levels begin at 
$\pm 1 \sigma$ and the contour interval is $1 \sigma$ (negative contours
are shown as dotted lines).  Although none of these 326 objects is
detected individually above the 3$\sigma$ level, average emission is detected 
for the ensemble at a significance of 3$\sigma$, and exhibits
a compact morphology centered on the mean stellar position. 
%In addition, the average
%image shows negative emission that resembles sidelobe
%features in the OVRO beam, suggesting that the average emission is
%point-like.
\label{fig:hist_o2}}
\end{figure}

Figure \ref{fig:hist_o2} shows the distribution of mm-wavelength
fluxes observed toward 326 $K$-band sources in the ONC, none of which are
detected individually above the 3$\sigma$ level.  
The 30 bright point sources visible in Figure \ref{fig:map_o2}
were removed using CLEAN before computing this histogram.
We also plot the flux distribution measured for all other pixels within the
unit gain contour (after subtraction of bright point sources).
The mean flux observed for the ensemble of 326 near-IR cluster members is 
0.267 mJy, and the standard deviation of the mean is $8.75 \times 10^{-5}$ Jy.
The flux distribution is thus biased to positive values with respect to the
noise distribution, and the significance of this bias is 3$\sigma$.
It appears
that while none of these $K$-band objects are detected individually 
in 3 mm continuum
emission above the 3$\sigma$ level, there may be weak emission below the
3$\sigma$ level from circumstellar disks.  

This positive bias is also illustrated in the right panel of
Figure \ref{fig:hist_o2}, which shows an
average image of the 3 mm flux observed toward $K$-band sources,
obtained by averaging $10'' \times 10''$ images centered around each object.
We CLEANed this average image in order to remove negative features
associated with sidelobe emission.  However, some negative features
remain, presumably as a result of residual calibration errors which prevent
perfect CLEANing.

The ``average'' disk is detected at a significance of $\sim 3\sigma$, and
is centered on the mean position of $K$-band sources (within positional
uncertainties).  In addition, the FWHM of the emission in Figure 
\ref{fig:hist_o2}, $2\rlap{.}''5 \times 1\rlap{.}''3$, is comparable to
the FWHM of the beam, suggesting that the 
positive bias observed in Figure 
\ref{fig:hist_o2} represents underlying
weak mm-wavelength emission from point sources.
We conclude that the 
average flux is dominated by small-scale emission such as that
expected for circumstellar disks.  
%However, the average image is slightly
%extended with respect to the beam (which has a FWHM of 
%$1\rlap{.}''9 \times 1\rlap{.}''5$), suggesting that larger structures
%(of diameter $\sim 1200$ AU) may also contribute to the average image.
%Thus we conclude that the 

Because the sensitivity of previous radio-wavelength surveys
\citep{FELLI+93a,FELLI+93b} is comparable to this average flux, we can
not rule out some contribution of free-free emission to the mean flux.
While the average flux arising from cool dust emission may therefore be 
somewhat
lower than 0.27 mJy,  we ignore this effect here, and await
more sensitive long-wavelength observations to provide better
constraints on low levels of free-free emission.
For the 326 known $K$-band sources, which predominantly
have stellar masses between
0.1 and 1 M$_{\odot}$ \citep{HC00}, our assumptions in 
Equation \ref{eq:mass_o2}
should be valid, and the conversion from 3 mm flux into mass should be reliable
(modulo the uncertainties concerning free-free emission discussed above).  
The mean
circumstellar mass among the 326 low-mass $K$-band sources within the unit
gain contour of our OVRO mosaic is $0.0055 \pm 0.0018$ M$_{\odot}$.

\section{Discussion \label{sec:disc_o2}}

\subsection{Frequency of Massive Disks \label{sec:disc_mass}}
We detected 30 objects in the ONC in 3 mm continuum emission, 10 of
which correspond to near-IR cluster members.  As discussed above,
for objects detected only in 3 mm emission, we can not determine whether
the dust is distributed in disks, envelopes, or combinations of the two.
Furthermore, for massive stars the geometry and total mass of circumstellar
material are uncertain since the dust may be substantially hotter than we
assume above.  Thus, we exclude the BN Object and $\theta^1$OriA from the list
of near-IR+mm detected objects when discussing massive disks.

Since our 3$\sigma$ detection threshold for 3 mm emission corresponds to a mass
of $\sim 0.1$ M$_{\odot}$ (if 100\% of the observed flux arises from dust
emission), in this section we will consider the statistics of disks more
massive than 0.1 M$_{\odot}$.  For the eight presumably low-mass mm+IR 
detected sources, the 3 mm emission from two objects (HC 322 and HC 323) is 
dominated by free-free emission, leaving only six massive disk candidates.  
Furthermore, the 1$\sigma$ uncertainty on the free-free emission contribution 
to the 3 mm flux for HC 241 cannot exclude a disk mass $\la 0.1$ M$_{\odot}$.
Thus, it appears that at most 6/336  ONC cluster members  ($\le 2\%$)  
possess massive circumstellar disks.

The large dust mass inferred for some of
these objects is higher than typically found for circumstellar disks 
\citep[e.g.,][]{BECKWITH+90}.  This may imply that some of these 
young stars are
surrounded by flattened disk+envelope structures, such as those inferred
around some Class I objects \citep{EISNER+05b}.  
However, we note that the two highest-mass disks are found in crowded regions,
and thus the large inferred masses may be contaminated by sidelobe 
emission from
nearby bright sources (Figure \ref{fig:detections}; \S \ref{sec:detections}).

Our observations did not find large millimeter fluxes toward the positions
of most near-IR cluster members in the ONC: $\le 2\%$ 
of the near-IR cluster
members have disk masses higher than $\sim 0.1$ M$_{\odot}$.
Although K-L color excesses of ONC members suggest that 
$80 \pm 7\%$ of the stars have circumstellar disks \citep{LADA+00}, these
two estimates of the disk fraction are not contradictory,
since the near-IR emission probes trace
material ($\sim 10^{-6}$ M$_{\odot}$) within $\sim 0.1$ AU of the star, while
the millimeter emission traces massive ($\ga 0.1$ M$_{\odot}$) outer
circumstellar material.  Furthermore, when treated as an ensemble, millimeter
emission is detected from near-IR objects in the mean, implying an
average disk mass of $\sim 0.005$ M$_{\odot}$.  

Based on our analysis of the ensemble of non-detected sources, it appears
that most stars in the ONC
likely possess disks with masses on the order of 0.005 M$_{\odot}$.
Thus, the ``average disk'' in this region is comparable to the 
minimum mass solar nebula \citep[$\sim 0.01$ M$_{\odot}$;][]{WEID+77},
indicating that most stars in the Trapezium region may be capable of 
forming solar systems like our own.  Since most stars in the Galaxy,
including our sun, probably formed in regions like Orion, it seems that
the capability to form analogs of our solar system is fairly common.

Our survey covered a larger field 
than the recent observations of \citet{WAW05},
but with a lower mass sensitivity.  Because \citet{WAW05} observed at 
850 $\mu$m, even though their flux sensitivity (2.7 mJy) was poorer than ours
($\sim 1.75$ mJy), the steep spectral dependence of dust emission
(Equation \ref{eq:dust}) means that their observations were sensitive
to smaller dust masses.  Thus, \citet{WAW05} detected three additional objects
(OW 170-337, OW 171-334, and OW 171-340) with circumstellar masses of 
$\sim 0.01$ M$_{\odot}$, which were below our detection threshold.  
However, the larger area covered by our mosaic allowed 
us to detect emission from massive disks around four proplyds
(OW 177-341, OW 158-327, OW 158-314, and OW 182-413) that were
outside of the \citet{WAW05} survey area.

\subsection{Dependence of Disk Properties on Environment \label{sec:disc_env}}
It has been suggested that circumstellar disks in clustered environments
may be truncated due to close encounters with massive stars resulting
in either tidal stripping or photo-evaporation of outer disk material
\citep[e.g.,][]{SC01}.
Evidence of this hypothesis has recently been inferred from the
observation that brown dwarfs appear more common in the ONC than in the
lower-stellar-density Taurus region \citep[e.g.,][]{KB03}, although this
result and its interpretation remain controversial \citep[e.g.,][]{GUIEU+05}.
While the observations presented
here do not have sufficient angular resolution to spatially resolve the outer
disk radius, we can test the truncation hypothesis by comparing
the disk mass distribution in the ONC and Taurus since truncated disks will be
less massive than un-truncated disks.

As discussed in \S \ref{sec:disc_mass}, we
detected disks more massive than $\sim 0.1$ M$_{\odot}$ 
around $\le 2\%$ of low-mass ONC cluster members.
For comparison,  $\sim 1.2\%$ of Taurus stars possess such
massive disks \citep{BECKWITH+90,OB95,AW05}.  The average disk masses in the
two regions are also comparable.
Thus, the fraction of massive 
disks in Taurus and Orion appears similar, arguing against the hypothesis 
that massive disks are truncated 
through close encounters with high-mass cluster members.  
We caution, however, that if dust temperatures in the ONC are
systematically higher than in Taurus (Equation \ref{eq:mass_o2};
\S \ref{sec:masses}), then the
percentage of high-mass disks in the ONC would be reduced
relative to Taurus.

Environmental effects on massive disks can also be investigated through
the dependence of disk properties on cluster radius.  
We therefore consider the positions of the six
massive disks detected in our observations (i.e., the subset of the 10
near-IR+mm detected objects that are not massive stars and are not dominated
by free-free emission; see \S \ref{sec:disc_mass}) relative to the
cluster center, which we define to lie
roughly in the middle of the four bright Trapezium
stars at $(\alpha,\delta)_{\rm J2000} = (5^{\rm h}35^{\rm m}16.34^{\rm s},
-5^{\circ}23'15\rlap{.}''6)$.   Within $10''$ of the cluster center, 
8\% of stars (1/13) possess massive disks, while this percentage falls to
4\% (1/25) between $10''$ and $20''$, and never higher than 5\% in successive
annuli throughout the remainder of the unit gain contour (4/290 sources
with radii $\ge 20''$).  While projection effects may complicate the
interpretation of this apparent trend, the marginally higher percentage 
of massive disks close to the massive Trapezium stars provides some further
support against the hypothesis of disk truncation through close encounters.

Although the distribution of massive disks in the ONC can not be
distinguished from that of Taurus, previous observations of the
disk mass distributions in the rich clusters NGC 2024 \citep{EC03} and 
IC 348 \citep{CARPENTER02} showed a statistically significant difference
from Taurus.  Based on these earlier results, we argued previously that 
the physical conditions in rich clusters might be responsible for observed
differences between clustered and isolated environments.  
%While our new
%results for the ONC argue that some clusters may be more similar to Taurus,
More sensitive interferometric observations are necessary to detect millimeter 
emission from a larger sample of disks, to place more definitive constraints 
on correlations of disk mass with environment.

\subsection{Disk Evolution \label{sec:disc_evol}}
We compare our results for the ONC with analogous
3 mm continuum surveys of NGC 2024 \citep{EC03} and 
IC 348 \citep{CARPENTER02}. 
NGC 2024 and IC 348, are somewhat less dense than the ONC, but still
each contain on the order of $300$ stars \citep{LADA+91,HERBIG98}.
In addition, spectroscopically-determined masses in the ONC 
\citep{HILLENBRAND97} and IC 348 \citep{LUHMAN99}, and estimated
masses from color-magnitude diagrams in NGC 2024 \citep{MEYER96,EC03} 
indicate similar stellar mass ranges in the three clusters,
although the spectral types of the most massive stars are somewhat cooler
in NGC 2024 \citep[$\sim$O8;][]{BIK+03} and IC 348 
\citep[B5;][]{LUHMAN+98} than in Orion.  
Comparison of the disk mass distributions can provide insights into
disk evolution, since the three clusters have apparently different ages.
Ages computed by comparing cluster member positions in an HR diagram
to the predictions
of pre-main-sequence evolutionary tracks \citep[using the same models for each
cluster;][]{DM97}, as well as the degree of embeddedness for each cluster
indicate that NGC 2024 is probably younger then the ONC, which is in turn 
younger than IC 348
\citep{MEYER96,ALI96,HILLENBRAND97,LUHMAN+98,LUHMAN99}.

\begin{figure}
\plotone{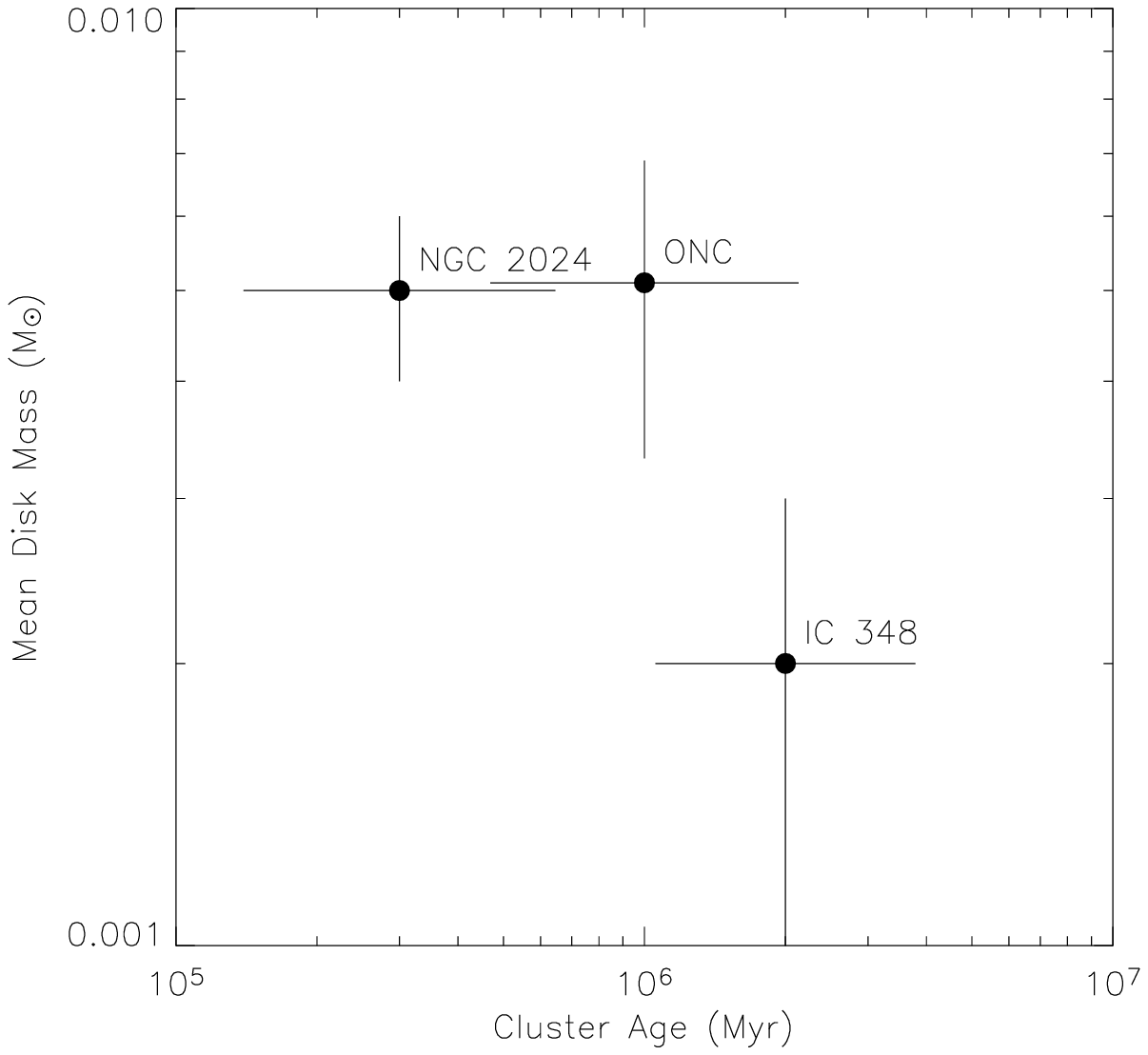}
\caption[Disk mass, as a function of cluster age]
{Average disk mass as a function of age for the NGC 2024, ONC, and
IC 348 clusters.  The disk masses are taken from this work, \citet{EC03},
and \citet{CARPENTER02}, and estimated cluster ages and uncertainties are from
\citet{MEYER96}, \citet{ALI96}, \citet{HILLENBRAND97}, \citet{LUHMAN+98},
and \citet{LUHMAN99}.
\label{fig:evol}}
\end{figure}

The average disk masses for ``typical'' stars in the three regions (i.e.,
the vast majority of cluster members not detected individually in 3 mm
emission)
is plotted as a function of cluster age in Figure \ref{fig:evol}.
In NGC 2024, the mean disk mass is $0.005 \pm 0.001$ M$_{\odot}$
\citep{EC03}, compared to 
$0.0055 \pm 0.0018$ in the ONC, and 
$0.002 \pm 0.001$ M$_{\odot}$ in IC 348 \citep{CARPENTER02}.  
In addition, the fraction of objects
detected in millimeter emission (without near-IR counterparts) is higher
in NGC 2024 ($\sim 6\%$) and the ONC ($\la 10\%$) than in IC 348 (0\%). 
In fact, since the observations of the ONC had a poorer 3$\sigma$ mass 
sensitivity for individual sources, 0.1 M$_{\odot}$ versus 0.035 M$_{\odot}$
in NGC 2024 \citep{EC03} and 0.025 M$_{\odot}$ in IC 348 \citep{CARPENTER02},
the fraction of structures in the ONC with comparable masses to those
detected in NGC 2024 may be somewhat higher.  Assuming that the differences
between NGC 2024, the ONC, and IC 348 are due to temporal evolution, these
observations indicate that massive disks/envelopes may dissipate on timescales
$\la 2$ Myr, and that the average disk mass may
decrease by a factor of $2.5 \pm 1.3$ between $\sim 0.3$ and 
2 Myr.  

It is important to keep in mind that the total millimeter emission is 
sensitive to dust grain properties in addition to total dust mass.  For
example, a given mass of dust grains larger than the observing wavelength 
emits less radiation than the same mass of small dust particles,
and the millimeter flux therefore depends on 
dust grain properties in addition to the total mass.  Furthermore, the
presence of hotter stars in the ONC (relative to NGC 2024 and IC 348) 
may potentially lead to systematically higher disk temperatures, which would
lead to smaller inferred masses 
(Equation \ref{eq:mass_o2}): if $T_{\rm dust}=50$
K, disk masses would be reduced by a factor of 3.3.
Thus, observed evolution in the millimeter flux may indicate that one
or more of the assumed quantities in Equation \ref{eq:mass_o2}
(e.g., temperature or opacity) is different in the three regions. 

Regardless of the underlying factors, our measurements 
provide a 2$\sigma$ suggestion of evolution between $\sim 0.3$ and 2 Myr.  
Future,
more sensitive measurements of the disk mass distributions in larger
numbers of clusters will decrease
the uncertainties in Figure \ref{fig:evol}, enabling more concrete
constraints on the evolutionary timescales of massive disks.
Moreover, follow-up observations at multiple wavelengths will begin 
to break degeneracies between dust grain properties and temperatures,
and total disk masses.

\section{Conclusions \label{sec:conc}}
We observed a $2\rlap{.}'5 \times 2\rlap{.}'5$ region of the Orion
Nebula cluster in $\lambda$3 mm continuum emission with the Owens Valley 
Millimeter Array.  The mosaic encompassed 336 young stars in the vicinity
of the massive Trapezium stars, and constrained the disk mass distribution
for this large number of cluster members.

We detected 30 objects in 3 mm continuum emission above the  3$\sigma$ limit,
10 of which correspond with near-IR cluster members.  Six of these, in
turn, also correspond with optically-detected proplyds.  
Comparison of our measured fluxes with longer wavelength observations
enabled rough separation of emission due to dust and that due to thermal
free-free emission, and we found that the 3 mm emission toward 8 
objects likely arises, in part, from dust.

We argued that with the exception of the massive stars $\theta^1$OriA and
the BN object, sources detected in both 3 mm and near-IR emission are probably
young stars surrounded by disks, and we computed circumstellar masses
of $0.13$-$0.39$ M$_{\odot}$ based on observed 3 mm fluxes (these
masses are uncertain by at least a factor of three due to uncertainties
in converting flux into mass).
Since the vast majority ($\ga 98\%$) of near-IR cluster members do not 
possess disks more massive than $\sim 0.1$ M$_{\odot}$, we placed constraints 
on lower-mass disks by considering 
the ensemble of 326 non-detected, predominantly low-mass stars.  For the
ensemble, we computed an average disk mass of 0.005 M$_{\odot}$, which
has a statistical significance of 3$\sigma$ (although the absolute
value of the mass contains larger uncertainties related to
converting flux into mass).

The average disk in the ONC is thus comparable to the minimum mass solar
nebula, suggesting that most young ($\la 1$ Myr)
stars in richly clustered environments (and
by extension, most stars in the Galaxy) are surrounded by sufficient 
circumstellar masses to form solar systems like our own.  Furthermore,
we found that the frequency of massive disks in the ONC appears similar
to that in Taurus, perhaps refuting suggestions that disks in Orion 
are truncated due to close encounters with the massive Trapezium stars.
Finally, we compared the disk mass distributions in three clusters
of different ages to begin to constrain the evolutionary timescales 
of massive disks.  Although substantial uncertainties remain, it
appears that massive disks may evolve significantly on 1-2 Myr timescales.

\noindent{\bf Acknowledgments.}  JAE is currently supported by a 
Miller Research Fellowship, and acknowledges past support from a
Michelson Graduate Research Fellowship.
JMC acknowledges support from 
the Owens Valley Radio Observatory, which is supported by the National Science
Foundation through grant AST-9981546.  The authors also wish to thank
John Bally for providing HST images of the Orion proplyds.
%This publication makes use of data products from the Two Micron
%All Sky Survey, which is a joint project of the University of Massachusetts
%and the Infrared Processing and Analysis Center, funded by the National
%Aeronautics and Space Administration and the National Science Foundation.
%2MASS science data and information services were provided by the Infrared
%Science Archive (IRSA) at IPAC.

%\epsscale{0.7}
%\begin{figure}
%\plotone{figs/orion_vlt_jhk.epsi}
%\caption[$JHK$ mosaic of the Orion Nebula cluster]{A $7' \times 7'$
%$JHK$ mosaic of the Orion Nebula cluster from VLT/ISAAC (courtesy of
%Mark McCaughrean and the European Southern Observatory).
%The bright, hot Trapezium stars are seen toward the center of the image,
%and hundreds of low-mass stars are seen in the central few arcminutes. 
%\label{fig:onc_jhk}}
%\end{figure}

\epsscale{0.7}

\epsscale{0.8}

\epsscale{0.7}

\epsscale{0.5}

\epsscale{0.35}

\epsscale{1.0}

\epsscale{0.8}

\clearpage
\begin{deluxetable}{lcccccccc}
\tabletypesize{\footnotesize}
%\tabletypesize{\small}
\tablewidth{0pt}
\tablecaption{Sources detected in $\lambda$3 mm continuum with OVRO 
\label{tab:detections}}
\tablehead{\colhead{ID} & \colhead{$\alpha$ (J2000)} & 
\colhead{$\delta$ (J2000)} & \colhead{$S_{\nu}$ (mJy)} & 
\colhead{$S_{\rm \nu, dust}$ (mJy)} & \colhead{$M_{\rm circumstellar}$ 
(M$_{\odot}$)} & 
\colhead{Other IDs}}
\startdata
\multicolumn{7}{c}{OVRO $>3\sigma$ detections with near-IR counterparts} \\
\hline
HC178$^{\ast}$ & 
5:35:13.54 & -5:23:59.4 & $15.55 \pm 2.2$ & & $0.32 \pm 0.05$ & \\
HC192$^{\ast}$ & 
5:35:13.57 & -5:23:55.2 & $18.92 \pm 2.2$ & & $0.39 \pm 0.05$ & \\
HC241 & 5:35:17.69 & -5:23:41.4 & $16.67 \pm 1.9$ & $7.40 \pm 5.5$ & 
$0.15 \pm 0.11$ & OW-177-341,F93-1 \\
HC282 & 5:35:18.85 & -5:23:28.6 & $6.50 \pm 1.5$ & & $0.13 \pm 0.03$ \\
HC287 & 5:35:15.82 & -5:23:26.6 & $18.59 \pm 1.9$  & $10.80 \pm 4.1$ &
$0.22 \pm 0.09$ & OW-158-327,F93-13 \\
HC322 & 5:35:16.30 & -5:23:16.6 & $11.30 \pm 1.8$ & $2.50 \pm 2.5$ & 
$0.05 \pm 0.05$ & OW-163-317,F93-7 \\
HC323 & 5:35:16.74 & -5:23:16.6 & $19.12 \pm 1.8$ & $0.00 \pm 5.5$ &
$0.00 \pm 0.10$ & $\theta^1$OriG,OW-167-317,F93-6 \\
HC336$^{\dag}$ & 5:35:15.81 & -5:23:14.6 & $19.70 \pm 1.8$ & $<15.80$ &
$<0.33$ & $\theta^1$OriA,OW-158-314,F93-12 \\
HC705$^{\dag}$ & 5:35:14.11 & -5:22:22.9 & $66.58 \pm 2.1$ & $60.90 \pm 4.1$ & 
$1.26 \pm 0.09$ & BN Object, F93-B \\
HC774 & 5:35:18.24 & -5:24:13.1 & $6.31 \pm 1.5$ & & $0.13 \pm 0.03$ &
OW-182-413,F93-O \\
\hline
\multicolumn{7}{c}{OVRO $>5\sigma$ detections without near-IR counterparts} \\
\hline
MM1 & 05 35 14.12 &  -05 22 04.6  &   $9.99 \pm 1.8$ &  & $0.21 \pm 0.04$ & \\
MM2 & 05 35 14.56  &  -05 22 19.6   & $9.67 \pm 1.9$ &  & $0.20 \pm 0.04$ & \\
MM3$^{\ast,\dag}$ & 05 35 14.53  &  -05 22 31.1 & 
$70.45 \pm 2.0$  & & $1.45 \pm 0.04$ & 
IRc2,F93-I \\
MM4$^{\ast}$ & 
05 35 14.86  &  -05 22 35.6   & $14.91 \pm 1.9$ & & $0.31 \pm 0.04$ & \\
MM5$^{\ast}$ & 
05 35 15.06  &  -05 22 03.1   & $26.03 \pm 2.0$ & & $0.54 \pm 0.04$ &  \\
MM6$^{\ast}$ & 
05 35 15.30  &  -05 22 05.1   & $26.80 \pm 1.9$ & & $0.55 \pm 0.04$ &  \\
MM7 & 05 35 13.72  &  -05 23 35.6   & $36.10 \pm 2.3$ &  & $0.74 \pm 0.05$ & \\
MM8$^{\ast}$ & 
05 35 13.69  &  -05 23 46.6   & $28.75 \pm 1.9$ &  & $0.59 \pm 0.04$ & \\
MM9$^{\ast}$ & 
05 35 13.72 &   -05 23 50.6   & $14.51 \pm 2.0$ &  & $0.30 \pm 0.04$ & \\
MM10$^{\ast}$ & 
05 35 14.03 &   -05 23 45.6   & $26.79 \pm 2.0$ & & $0.55 \pm 0.04$ & \\
MM11$^{\ast}$ & 
05 35 13.56 &   -05 24 02.6   & $15.43 \pm 2.3$ & & $0.32 \pm 0.05$ & \\
MM12$^{\ast}$ & 
05 35 13.49 &   -05 24 10.6   & $40.82 \pm 2.3$ & & $0.84 \pm 0.05$ & \\
MM13$^{\ast}$ & 
05 35 13.76 &   -05 24 08.1   & $36.92 \pm 2.3$ & & $0.76 \pm 0.05$ & \\
MM14$^{\ast}$ & 
05 35 13.92 &   -05 24 09.1   & $18.35 \pm 1.9$ & & $0.38 \pm 0.04$ & \\
MM15 & 05 35 13.96 &   -05 23 57.1   & $11.72 \pm 1.9$ & & $0.24 \pm 0.04$ & \\
MM16 & 05 35 14.36 &   -05 23 54.6   & $16.02 \pm 2.0$ & & $0.33 \pm 0.04$ & \\
MM17 & 05 35 15.16 &   -05 23 40.6   & $15.31 \pm 2.0$ & & $0.32 \pm 0.04$ & \\
MM18 & 
05 35 15.36 &   -05 23 05.1   & $12.52 \pm 2.3$ & & $0.26 \pm 0.05$ & \\
MM19$^{\ast}$ & 
05 35 15.93 &   -05 23 45.6   & $10.31 \pm 2.0$ & & $0.21 \pm 0.04$ & \\
MM20$^{\ast}$ & 
05 35 15.93 &   -05 23 50.1   & $11.94 \pm 2.3$ & & $0.25 \pm 0.05$ & \\
\enddata
\tablecomments{$S_{\rm \nu,dust}$ is the component of the observed
3 mm emission due to cool dust, determined from a fit to long-wavelength
fluxes of a model including thermal free-free emission as well as cool dust 
emission (see Figure \ref{fig:seds}).  All quoted uncertainties are
1$\sigma$, and include only the RMS uncertainties in the measured fluxes;
systematic uncertainties associated with the conversion from flux to
mass are not included. $^{\ast}$--These objects are located in crowded
regions, and thus the observed flux may be contaminated by sidelobe emission
from bright, nearby sources. $^{\dag}$--$\theta^1$OriA, BN, and IRc2 
are massive stars,
and since the dust around these objects may be hotter than the 20 K used
to compute $M_{\rm circumstellar}$, the 
circumstellar masses are probably overestimated. }
\end{deluxetable}

\end{document}